\documentclass[12pt,a4paper,rotating]{article}
\usepackage{a4p}
\usepackage{times,psmath}
\usepackage{cite,mcite}
\usepackage{graphicx}
\usepackage{physics}
\usepackage{ams_title,ifthen,Lep}
\journalname{Phys.~Lett.~B}
\journal   %     for journal version
\newlength{\capindent}
\newlength{\capwidth}
\newlength{\figwidth}
\newlength{\figheight}
\newcommand{\icaption}[2][!*!,!]{\hspace*{\capindent}%
  \begin{minipage}{\capwidth}
    \ifthenelse{\equal{#1}{!*!,!}}%
      {\caption{#2}}%
      {\caption[#1]{#2}}
  \end{minipage}}
\newcommand{\+}{$\,$}
\newcommand{\LL}[1]{\smash{\lower 1.5ex \hbox{#1}}} %for table
\newcommand{\Hethree}{\ensuremath{^3\kern-0.05em\mathrm{He}}}
\newcommand{\Hefour}{\ensuremath{^4\kern-0.05em\mathrm{He}}}
\newcommand{\diff}[2]{\frac{\mathrm{d}#1}{\mathrm {d}#2}}
\newcommand{\STRUT}{\rule[-1ex]{0ex}{3.5ex}}

\begin{document}
\begin{titlepage}

\title{Cosmic-ray positron fraction measurement\\from 1 to 30\+\GeV{} with AMS-01}
\author{The AMS-01 Collaboration}

% make the title
%\maketitle               

% do the abstract
\begin{abstract}
A measurement of the cosmic ray positron fraction $e^+/(e^++e^-)$ in
the energy range of \mbox{1--30\+\GeV{}} is presented. The measurement
is based on data taken by the AMS-01 experiment during its 10 day
Space Shuttle flight in June 1998. A proton background suppression on
the order of $10^{6}$ is reached by identifying converted
bremsstrahlung photons emitted from positrons.
\end{abstract}
\submitted
\end{titlepage}

%----------------------------------------------------------------------
% SECTION I: Introduction
%----------------------------------------------------------------------
\section{Introduction}
Over the past decades cosmic ray physics has joined astronomy as a
means to gather information about the surrounding universe. Of the few
particles that are stable and thus able to cross vast interstellar
distances, electrons and positrons are of particular
interest.

Electrons are believed to be accelerated in shock waves following
supernova explosions. Their spectrum is subsequently altered by
inverse Compton scattering off cosmic microwave background photons,
synchrotron radiation due to the galactic magnetic field,
bremsstrahlung processes in the interstellar medium and modulation in
the solar magnetosphere. Thus they serve as an important probe of
cosmic ray propagation models. On the other hand, positrons are
produced secondarily in the decay cascades of $\pi^+$, which are
created in hadronic interactions of cosmic ray protons with the
interstellar medium. This yields an $e^+/e^-$ ratio of roughly 10\+\%.

In addition to these classical sources, positrons may also originate
from more exotic ones. Among the most important unsolved questions in
modern cosmology is the nature of dark matter. Based on observations
of the cosmic microwave background, supernovae of type IA and galaxy
clustering, among others, the standard model of cosmology now contains
a density of non-luminous matter exceeding that of baryonic matter by
almost a factor of five~\cite{eidelmann04}. The most promising
candidate for dark matter is a stable weakly interacting massive
particle predicted by certain supersymmetric extensions to the
standard model of particle physics~\cite{turner90} and called the
neutralino, $\chi$. Positrons and electrons will then be created in
equal numbers as stable decay products of particles stemming from
$\chi$-$\chi$ annihilations, for instance in the galactic halo. Such a
process would constitute a primary source of positrons. Therefore, a
measurement of the positron fraction is also motivated by the prospect
of indirect dark matter detection, especially if combined with other
sources of information, such as antiprotons, diffuse $\gamma$-rays or,
more challenging, antideuterons.

\section{The AMS-01 experiment}
As a predecessor to the Alpha Magnetic Spectrometer AMS-02, which is
to be operated on the International Space Station (ISS) for at least 3
years, the AMS-01 experiment was flown on the Space Shuttle {\it Discovery}
from June 2nd to 12th, 1998.

The AMS-01 experiment consisted of a cylindrical permanent magnet with
a bending power of 0.14\+Tm$^2$ and an acceptance of
0.82\+m$^2$sr. The magnet bore was covered at each of the upper and
lower ends with two orthogonal layers of scintillator paddles, forming
the time of flight system (TOF). This provided a fast trigger signal
as well as a measurement of velocity and charge number. The silicon
tracking device consisted of six layers of double-sided silicon strip
detectors mounted inside the magnet volume. Charged particle
trajectories were reconstructed with an accuracy of better than
20\+$\mu$m in the bending coordinate. The momentum resolution at
10\+\GeV{}/c was about 10\+\% for singly charged particles. The inner
magnet surface was lined with the scintillator panels of the
anticoincidence system serving as a veto counter against particles
traversing the magnet wall. Velocity measurements were augmented with
a two-layered aerogel
\v Cerenkov threshold counter (ATC) mounted underneath the lowest TOF
layer, allowing $e^{+}/p$ discrimination below 3\+\GeV{}/c. A low
energy particle shield covered the experiment to absorb particles
below 5\+\MeV{}, while a multi-layer insulation blanket served as a
protection against space debris and solar radiation. The radiation
thickness of all materials above the tracking device sums up to
18.2\+\% of a radiation length. Below the tracking device, not
including the Space Shuttle, the material sums up to 19.1\+\% of a
radiation length. A detailed description of the experiment is given
in~\cite{aguilar02a}.
Orbiting with an inclination of $\pm\,$50.7$^{\circ}$ at altitudes
between 320 and 390\+km, AMS-01 recorded $10^8$ events in 184
hours. During 4 days of the flight, the Space Shuttle was docked to
the Mir space station. Before and after docking, the Shuttle's yaw
axis (AMS z-axis) was kept pointing at 0$^{\circ}$, 20$^{\circ}$,
30$^{\circ}$, 45$^{\circ}$ and 180$^{\circ}$ with respect to the
zenith for several hours each. During docking its pointing varied
continuously between 40$^{\circ}$ and 145$^{\circ}$.

\section{Conversion of bremsstrahlung photons}
The main challenge of cosmic ray positron measurements is the
suppression of the vast proton background. As is known from previous
measurements~\mcite{aguilar02a,duvernois01a}, the flux of cosmic
ray protons exceeds that of positrons by a factor of $10^4$ in the
momentum range of 1--50\+\GeV{}/c. Hence, in order to keep the proton
contamination of positron samples below 1\+\%, a proton rejection of
$10^6$ has to be reached. Since the ATC subdetector of AMS-01 provided
a sufficient single track proton rejection only for energies below
3\+\GeV{}, a different approach has been chosen for this analysis. It
relies on the identification of bremsstrahlung emission through
photoconversion. Due to the inverse quadratic dependence on the
particle mass of the cross section, bremsstrahlung emission is
suppressed by a factor of more than $3\cdot 10^{6}$ for protons with
respect to positrons.
%% FIGURE fig:convScheme

\begin{figure}[hb]
\begin{center}
\includegraphics[width=10cm]{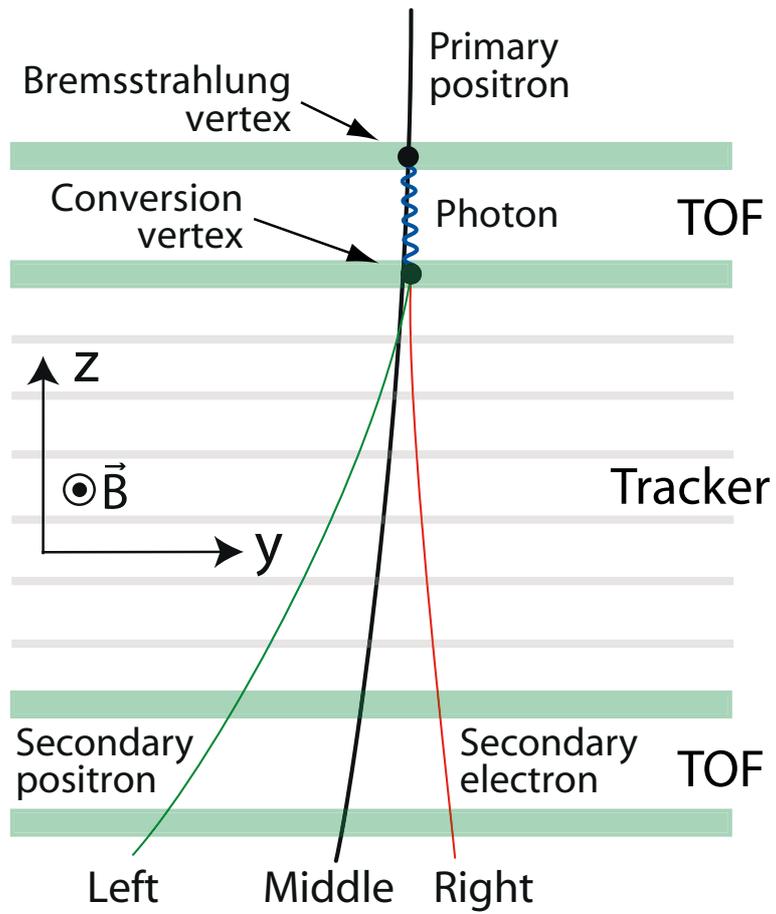}
\caption{\label{fig:convScheme}Schematic view of a converted bremsstrahlung event caused by a positron going top-down.}
\end{center} 
\end{figure}

Figure~\ref{fig:convScheme} shows the principle of a converted
bremsstrahlung event signature. A primary positron enters the
detector volume from above and emits a bremsstrahlung photon in the
first TOF scintillator layer. The photon then converts into an
electron-positron pair in the second TOF layer. Because of the low
fraction of momentum which is typically carried away by the photon,
the secondary particles have lower momenta than the primary.
Therefore, in the bending plane projection, the secondaries tend to
form the left and right tracks, while the primary remains in the
middle.

Both bremsstrahlung and photon conversion are closely related
electromagnetic processes whose energy and angular distributions can be
calculated with the Bethe-Heitler formalism. In the relativistic
limit, the angles of photon emission as well as the opening angles of
pair production show distributions with a most probable value of
$\theta_0\approx 1/\gamma$, $\gamma$ being the Lorentz factor of the
emitting particle or the electron-positron pair, respectively.  In the
GeV energy range, these values fall below the accuracy limit of the
track reconstruction induced by multiple scattering, and thus are
practically equal to zero.

The dominant backgrounds are caused by electrons with misreconstructed
momentum sign and by protons undergoing hadronic reactions in the
material distribution of the experiment. In the latter case, mesons
are produced that mimic the 3-track signature of converted
bremsstrahlung events. For example, in the reaction $pN\to
pN\pi^+\pi^-+X$, where $X$ are additional undetected particles, the
charged pions can be misidentified as an electron-positron
pair. Besides this, neutral pions produced in reactions of the type
$pN\to pN\pi^0+X$ decay into two photons, one of which may escape
undetected.  If the remaining photon converts, the conversion pair
will form a 3-track event together with the primary proton.  However,
the invariant masses of the mesons and the primary proton and photon
are typically at the scale of the pion mass, leading to significantly
larger emission angles.

\section{Event reconstruction}
In order to gain the highest possible selection efficiency, it is
mandatory to apply sophisticated track and vertex finding algorithms
which are particularly customized for the converted bremsstrahlung
event signature~\cite{jan}. To account for the asymmetric geometry of
the detector along its z-axis, the analysis is performed separately
for particles traversing the detector top-down {\it (downward)} and
bottom-up {\it (upward)}.

\subsection{Preselection}
As a first step, the hits in the silicon strip detectors of the
tracker are projected into the bending plane for clustering. For
further analysis, a minimum of 8 tracker clusters are required. Events
are selected in which at least two of the six layers of silicon
detectors signaled exactly three clusters {\it (triplets)\/}.

\subsection{Track finding}
Since particle tracks diverge in the magnetic field, the triplets are
required to have increasing cluster to cluster distances along the
z-axis in the flight direction. Assuming that three particles have
traversed the tracker, in events with three or more triplets the
clusters in the triplets can be directly assigned to a left, a middle
and a right track of minimum length. Starting with these track seeds,
further clusters on the other layers are gradually added to the
tracks. Layer by layer, a competition algorithm based on $\chi^2$
minimization builds the tracks and assigns as many clusters as
possible to them.

A generalized algorithm has been developed for the treatment of events
that feature only two tracker layers with exactly three clusters. It
is based on a combinatorial approach to the track finding problem and
has been employed in the analysis, thus improving the lepton
efficiency~\cite{henning}.

Subsequently, ambiguities in the x-coordinate, parallel to the
magnetic field, due to the clustering in the bending plane projection
only, have to be resolved. For this, a narrow corridor along the hits
in the TOF system is defined, and only tracker hits within this
corridor are retained. To each track, a series of helix fits is
applied, taking into account each combination of hits in any of the
track clusters. The final tracks are then defined by the combinations
with the lowest $\chi^2$.

\subsection{Vertice reconstruction}
Vertex reconstruction is based on back-propagation of the tracks
through the magnetic field using the functionality of the GEANT3
package~\cite{brun87}. The vertices of the left and right tracks are
determined by parallel back-propagation from the point of the first
tracker hit. The {\it conversion\/} vertex is then defined as the
barycenter of the track points at the z-coordinate of closest approach
of the tracks. In case the tracks intersect in the bending plane
projection, the intersection point is taken as the vertex with the
x-coordinate derived from geometrical interpolation.

The four-vector of the photon is reconstructed from the sum of those
of the left and right track. Then, using the same algorithm as
described above, the {\it bremsstrahlung\/} vertex of the photon and
the middle track is computed.

No requirements are placed on the location of the bremsstrahlung
vertex nor the conversion vertex.

\subsection{Reconstruction quality and Monte Carlo}
The quality of the reconstruction algorithms is verified with
$16.8\cdot 10^6$ electron and positron events from a complete Monte
Carlo simulation of the experiment using GEANT3. The momentum
resolution is approximately 13\+\% for the downward case and 14\+\%
for the upward case. This resolution is similar to that for single
track events in the energy range of 10\+\GeV{} and
above~\mcite{alcaraz99a,chang01}, where our reconstruction algorithms
have their peak sensitivity.

The properties of the bremsstrahlung photon can be particularly well
reconstructed. The momentum resolution of the photon is 8\+\%, while the
absolute direction error has a standard deviation of below 9\+mrad.

\section{Analysis}
\label{section:analysis}
Analysis and suppression of background mainly rely on the evaluation
of the topology and geometrical properties of the reconstructed
events, and are therefore based on data from the
tracker. Additionally, cuts on data from the TOF system are
applied. However, substantial parts of the analysis deal with measures
to account for the environmental circumstances under which the AMS-01
experiment was operated, especially the effect of the geomagnetic
field.

\subsection{Basic cuts}
\label{section:basiccuts}
Several cuts have to be applied to the data in order to suppress
misreconstructed events:
\medskip
\begin{itemize}
\item Track fits with resulting momenta lower than 100\+\MeV{}/c may
lead to misreconstruction. Events containing such tracks are thus
rejected.
\item Due to the deflection in the magnetic field, the
charge signs of the secondaries are exactly constrained and depend on
flight direction. The charge sum of the three tracks is required to be
$\pm$1.
\item With higher energies, the track momentum resolution and
the signal over background ratio deteriorate. Thus the total
reconstructed momentum must not exceed~50\+\GeV{}/c.
\end{itemize}
\medskip

The requirement for increasing cluster distances within the seed
triplets along the flight path largely distinguishes between downward
and upward going particles. To make sure the flight direction is
correctly recognized, timing information from the TOF system is
used. The time of passage in the individual scintillators is measured
with a resolution of 120\+ps~\cite{alvisi99a}. The flight time $t_f$
for downward and upward going particles is calculated according to
$t_f = (t_1+t_2)/2 - t_3$, where the $t_i$ denote the time of passage
measured in TOF layer~$i$ ($i$ is counted from top to bottom). Due to
high voltage failures in TOF layer~4~\cite{alvisi99a}, its timing
information is not used. The sign of $t_f$ depends on the flight
direction. Events are rejected for which $\left|t_f\right|$ is smaller
than 3.5\+ns or the sign of $t_f$ disagrees with the flight direction
given by the requirement of increasing cluster distances.

To make sure that there are three particles traversing the detector,
consistent with the signature of a converted bremsstrahlung photon, a
minimum average energy deposition of 5\+\MeV{} (equivalent to 2 MIPs)
is required in each of the last two TOF layers in the flight
direction.

Nuclei such as $He$ or $N$ have been observed to induce background
events through hadronic interactions. Such particles with $Z>1$
deposit significantly more energy in the subdetectors than singly
charged particles. The truncated mean of the energy depositions in the
TOF scintillators is calculated, and events are rejected with an
energy deposition of more than 10\+\MeV{}. Additionally, a cut is
applied to the mean of the three highest tracker hit amplitudes. By
these means events involving nuclei are entirely eliminated.

\subsection{Suppression of dominant background}
For the suppression of background, the fact is used that
bremsstrahlung and photon conversion imply small opening angles
of the particles at the vertices. In order to make these angles
independent of the frame of reference, the corresponding invariant
masses are calculated according to
\begin{equation}
m_{\,\mathrm{inv}}^2 = 2 \cdot E_1 \cdot E_2 \cdot \left( 1 - \cos\theta
\right),
\end{equation}
where $\theta$, $E_1$ and $E_2$ denote the opening angle and the
energies of the primary particle and the photon, or the conversion
pair, respectively.

The distribution of the invariant mass at the conversion vertex is
shown in Figure~\ref{fig:conversionInvMass}. For events with negative
charge, which represent a largely clean electron sample, it reveals a
narrow shape with a peak at zero, in agreement with Monte Carlo
results. For events with positive total charge, consisting of
positrons and background, the distribution also shows a peak at zero,
and an additional long tail towards higher invariant masses caused by
the proton background. The distributions of the invariant mass at the
bremsstrahlung vertex show similar behaviors. In order to
discriminate against background events, cuts are applied on the
invariant masses. The cuts are parameterized as ellipses in the
invariant mass plane, centered at zero, with half axes in units of the
standard deviations, $\sigma$, of the electron distribution from
data. Events outside the ellipses are rejected. In order to keep the
positron selection efficiency high, the cut values have been set to
2$\sigma$.
% FIGURE fig:conversionInvMass

\subsection{Geomagnetic cutoff}
\label{subsection:cutoff}
The spectra of cosmic rays are modulated by the geomagnetic
field. Depending on the incident direction and the geomagnetic
coordinates of the entry point into the magnetosphere, particles with
momenta below a certain cutoff are deflected by the geomagnetic field
and cannot reach the Earth's proximity. Hence, below geomagnetic
cutoff the particles detected by AMS-01 must originate from within the
magnetosphere. They were mostly produced as secondaries through
hadronic interactions and trapped on geomagnetic field lines.

To discriminate against these secondaries, particle trajectories were
individually traced back from their measured incident location, angle
and momentum through the geomagnetic field by numerical integration of
the equation of motion~\cite{flueckiger90a}. A particle was rejected as
a secondary if its trajectory once approached the surface of the
Earth, and thus originated from an interaction with the
atmosphere. Particles which did not reach a distance of 25 Earth radii
were considered as trapped and also rejected.

\section{Correction for irreducible background}
\label{section:correction}
\begin{figure}[!h]
\begin{center}
\includegraphics[width=14cm]{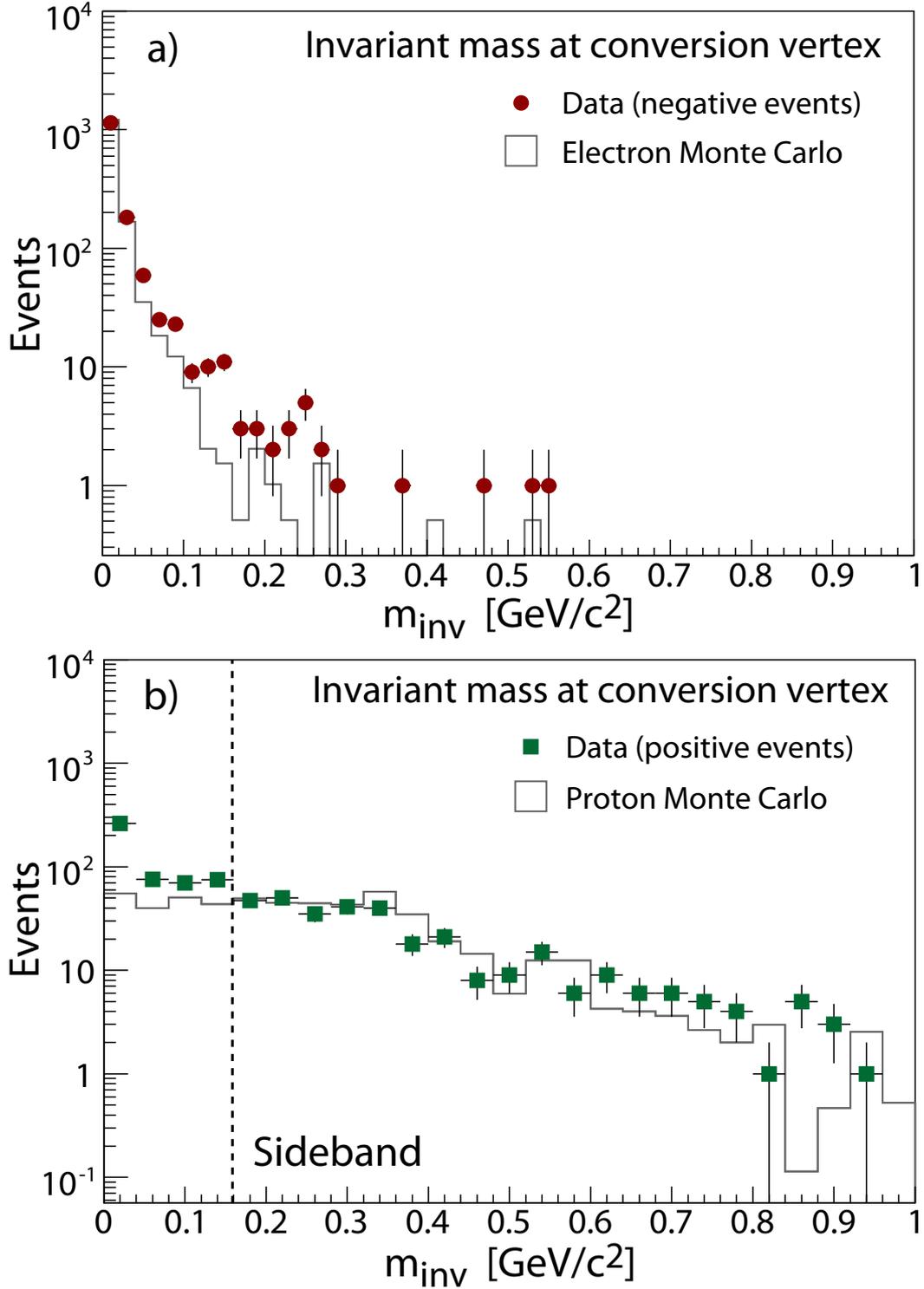}
\caption{\label{fig:conversionInvMass}a) Invariant mass distribution at the conversion
vertex for negatively charged data events (circles) and electron Monte
Carlo (histogram).  b) The same display for positively charged data
events (squares) and proton Monte Carlo (histogram). The proton Monte
Carlo distribution has been scaled to the data using the
sideband. Below the sideband threshold of 0.16\+\GeV{}/c$^2$, the excess
in the data due to the positron contribution is apparent.}
\end{center} 
\end{figure}

As can be seen in Figure~\ref{fig:conversionInvMass}, the invariant
mass distribution of protons does not vanish in the signal region. The
same applies to the background from misidentified
electrons. Consequently, a small fraction of background events will
not be rejected by the cut on the invariant masses. This remaining
irreducible background has to be corrected. This has been
accomplished using Monte Carlo simulations.

The approach used is to run the analysis on $16.5\cdot 10^7$ proton
and $9.4\cdot 10^6$ electron Monte Carlo events as if they were data,
determine the momentum distribution of particles that are
misidentified as positrons, and subtract these from the raw positron
counts obtained from data. However, such a comparison of Monte Carlo
and data requires the adjustment of several properties of the
simulated events. Particularly, they have not been affected by the
geomagnetic field.

As introduced in \S~\ref{subsection:cutoff}, the geomagnetic field
shields the Earth's vicinity from low energy particles. However, the
geomagnetic cutoff cannot be calculated individually for Monte Carlo
particles, since their four vector is not defined with respect to the
geomagnetic coordinates. To correct for the shielding effect, the
livetime function, described in \S~\ref{sec:livetime} is
used. The livetime function gives the effective measurement time as a
function of momentum for singly charged particles. Normalized to a
maximum value of 1 at highest momenta well above the cutoff, its value
at a given momentum denotes the probability for a particle to
penetrate the geomagnetic field. Hence, it serves as a weight for
distributions of any event variable from Monte Carlo, particularly for
the momentum distribution of background events from Monte Carlo. As
for the data, the livetime function has to be evaluated using the
reconstructed momentum, rather than the incident particle's simulated
momentum.

The incident momentum spectrum of the Monte Carlo particles follows a
distribution $\phi_{MC}(p) = p^{-1}$, which differs significantly from
the true spectrum. Since the event variables are correlated with the
incident momentum, they again have to be reweighted. Using the
parameterized fluxes $\phi_D(p)$ of protons~\cite{aguilar02a} and
electrons~\cite{alcaraz00a}, measured by AMS-01, the spectral
reweighting function is calculated as $w(p)=\phi_D(p)/\phi_{MC}(p)$.

The livetime function as well as the spectral reweighting function
correct for the shape of the momentum distribution of background events
calculated from Monte Carlo. Subsequently, since the latter function
does not conserve the integral, the background distributions need to
be scaled to the data.

Figure~\ref{fig:conversionInvMass}b illustrates the scaling of the
proton Monte Carlo to the data using the sidebands of the invariant
mass distributions. The sidebands are defined as the ranges of
invariant mass above certain thresholds in which the positron
contribution to the sample of positively charged events from data is
negligible. The thresholds are determined from the electron
distribution to be 0.16\+\GeV{}/c$^2$ at the conversion vertex and
0.2\+\GeV{}/c$^2$ at the bremsstrahlung vertex. Below the thresholds the
excess in the data due to the positron contribution is apparent.
% FIGURE fig:subtractedBackground

The correction due to electrons with misreconstructed charge sign is
calculated in a very similar way. The main difference is the fact
that the distributions originating from a given number of Monte Carlo
electrons are scaled directly to the electron candidate sample found
in the data.

\begin{figure}[!h]
\begin{center}
\includegraphics[width=14cm]{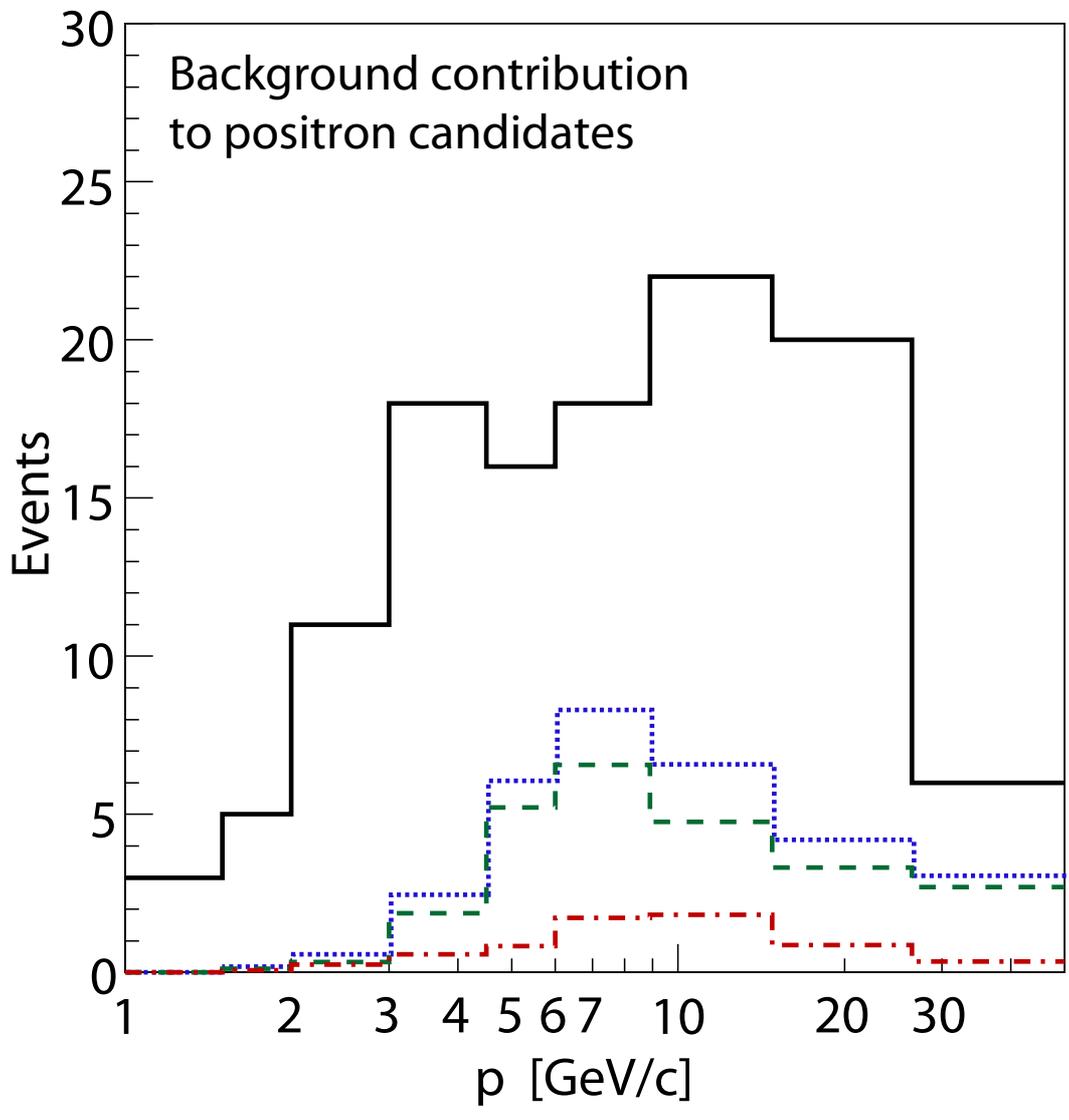}
\caption{\label{fig:subtractedBackground}Momentum distribution of the positron candidates 
including background (solid line) and the total estimated background (blue dotted line),
itemized into contributions from protons (green dashed line) and wrongly identified
electrons (red dash-dotted line).}
\end{center} 
\end{figure}

Using the scaling factors obtained with the above procedures, the
background contribution to the number of positron candidates is
calculated. Figure~\ref{fig:subtractedBackground} shows the total
background correction as a function of momentum, separately indicating
the contributions from protons and misidentified electrons. In total,
they amount to 24.9 and 6.5 events, respectively. The resulting
corrected lepton sample consists of 86 positrons and 1026 electrons.

\section{Positron fraction}
\label{section:fraction}
\begin{figure}[!h]
\begin{center}
\includegraphics[width=14cm]{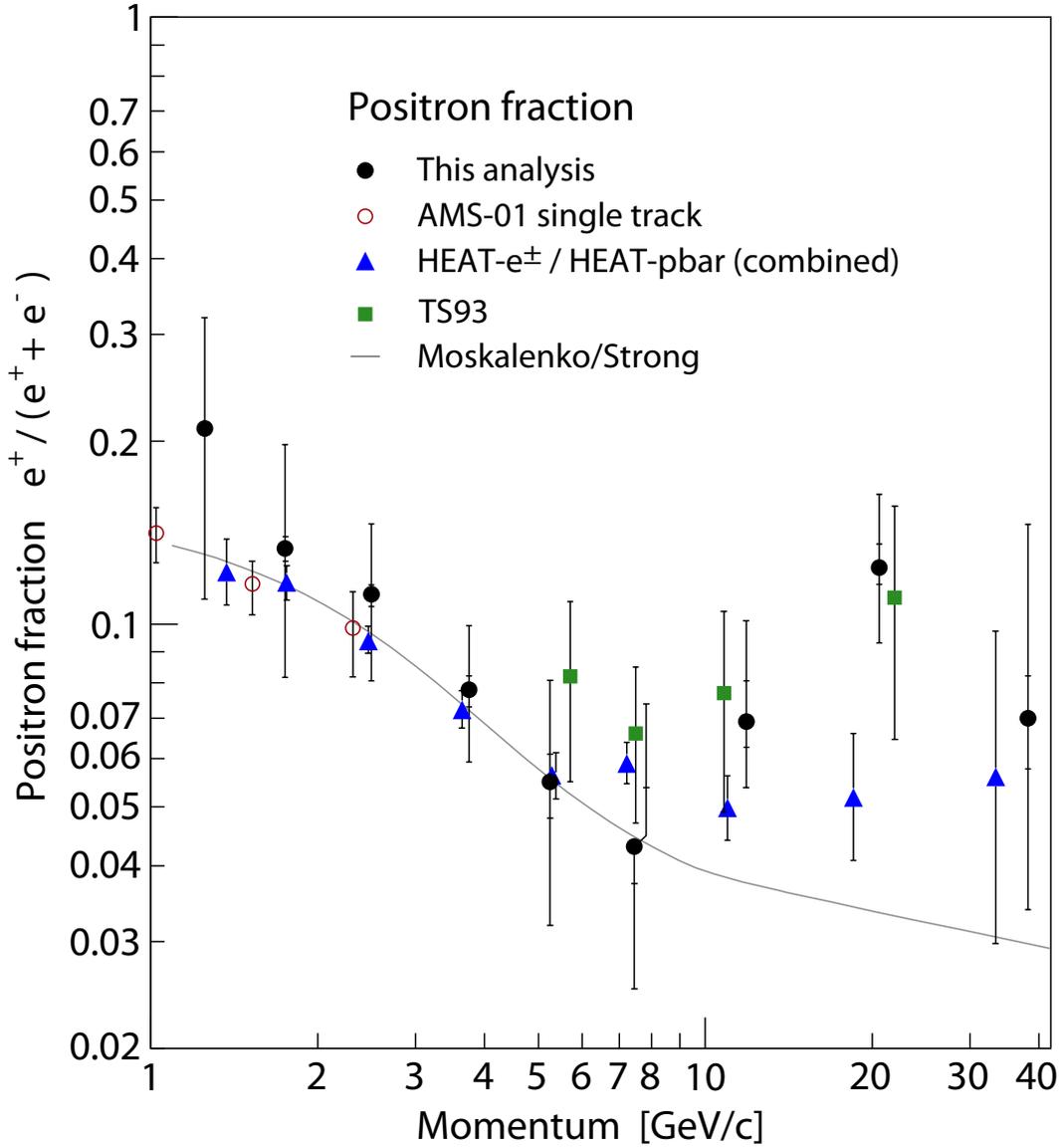}
\caption{\label{fig:fraction} The positron fraction $e^+/(e^{+}+e^-)$
measured in this analysis (filled circles), compared with earlier
results from AMS-01 (open circles)~\protect\cite{alcaraz00a},
TS93 (squares)~\protect\cite{golden96a}, the combined results from HEAT-$e^{\pm}$ and
HEAT-pbar (triangles)~\protect\cite{beatty04a}, together with a model
calculation for purely secondary positron production from
\protect\cite{moskalenko98a} (solid line). The total error is given
by the outer error bars, while the inner bars represent the systematic
contribution to the total error.}
\end{center} 
\end{figure}

The positron fraction $e^+/(e^{+}+e^-)$ is calculated from the
electron counts and corrected positron counts for each energy bin. It
is shown in Figure~\ref{fig:fraction} in comparison with earlier
results~\mcite{alcaraz00a,golden96a,beatty04a} and a
model calculation based on purely secondary positron
production~\cite{moskalenko98a}. Table
\ref{table:fraction} summarizes the results. The total errors are
clearly dominated by the contribution from statistical errors,
systematic errors play a lesser role. In the following, the
contributions to the error on the positron fraction are discussed.
%FIGURE fig:fraction

\subsection{Statistical errors}
Due to the complexity of the positron fraction computation, taking
into account two sources of background, and low statistics, a Bayesian
approach based on Monte Carlo simulation has been chosen for the
determination of the statistical errors~\cite{helene83a}. The aim is
to acquire the probability distribution of all possible values of the
positron fraction which can, superimposed on the background, lead to
the observed number of particle counts. From this distribution, the
confidence levels are derived by numerical integration.

In a first step, for a particular momentum bin, two random floating
point numbers are generated, following a uniform distribution and
representing the ``true'' numbers of electrons and
positrons. Subsequently, the background counts from Monte Carlo --
modulated with errors to account for their systematic uncertainty --
are added to the true number of positrons. Here, the scale factors
from background scaling have to be considered. The resulting numbers
of positively and negatively charged particles are modulated with
Poisson errors, thus become integers, and then represent the
``measured'' number of candidates including background. If these
numbers are exactly equal to the counts observed in the experiment,
the positron fraction calculated from the true numbers is accepted for
further analysis, and the above procedure is repeated.

The distribution of simulated positron fraction values is finally
parameterized and normalized to an integral of 1. Subsequently, by
repeated numerical integration, the smallest interval is found in
which the integral of the distribution equals 0.683, consequently
giving the lower and upper limit of the 1$\sigma$ Gaussian confidence
interval.

\subsection{Systematic errors}
In the positron fraction -- as a ratio of particle fluxes -- most sources
of systematic error, such as detector acceptance or trigger efficiency,
naturally cancel out. Hence, only sources of error which are asymmetric
with respect to the particle charge have to be considered.

Background correction is applied to the sample of positron candidates
only and is therefore a source of systematic error. To a certain
degree, the description of the experimental setup may be inaccurately
implemented in the Monte Carlo program. Furthermore, in contrast to
the production of charged pions, background processes involving
neutral pion production imply photoconversion with typically low
angles between tracks emerging from the vertices. Hence, the
distribution of invariant masses depends on the cross
sections of charged and neutral pion production. Possible inaccuracies
in the implementation of the cross sections in the Monte Carlo program
must therefore be considered.

The systematic error from background correction can be estimated by
evaluating the deviation of the scaled Monte Carlo background from the
data in the invariant mass plane. With a binning coarse enough to
flatten statistical fluctuations, the mean deviation outside the
signal region leads to a systematic error estimate of 20\+\% of the
background events. This value is then propagated to the positron
fraction for each momentum bin.

As a consequence of the East-West effect~\cite{johnson34a}, in
combination with the asymmetric layout of the AMS-01 tracker, the
product of the detector acceptance times the livetime as functions of
the particles' incident direction may vary for positrons and
electrons. Even though no deviation of their average livetimes is
apparent (see \S~\ref{sec:livetime}), we account for
this effect with a second contribution to the systematic error of the
positron fraction. It is estimated from the mean variation of the
difference in livetime of positrons and electrons over the detector
acceptance.  After propagation to the positron fraction, the
systematic error due to the East-West effect is below 10\+\% for
all momentum bins, except for the highest momenta above 26.5\+\GeV{}, where
it amounts to approximately 10\+\% of the positron fraction value.

\section{Flux calculation}
\label{section:fluxcalculation}
As a crosscheck to the measurement of the positron fraction, presented
above, the absolute incident fluxes of electrons and positrons are
calculated. The electron flux is then compared to measurements by
other experiments and the results obtained previously by AMS-01.

One can calculate the differential flux for a given momentum bin $p$
of width $\Delta{}p$ from the measured particle count
$N(p,\theta,\phi)$ in this bin, the detector acceptance
$A(p,\theta,\phi)$, and the livetime $T(p,\theta,\phi)$, as follows:
\begin{equation}\label{accurate_flux}
\diff{\Phi(p,\theta,\phi)}{p}=\frac{N(p,\theta,\phi)}{A(p,\theta,\phi)\cdot{}T(p,\theta,\phi)\cdot{}\Delta{}p}\, .
\end{equation}
By the term {\it livetime}, we mean the effective amount of time
during which cosmic ray particles coming from outer space have the
opportunity to reach the detector. If -- as is the case with the
AMS-01 downward flux -- the livetime is only weakly depending on the
direction, the angular distribution of the particle count will follow
that of the acceptance. Then, one can approximate
(\ref{accurate_flux}) to become
\begin{equation}
\diff{\Phi(p)}{p}=\frac{N(p)}{A(p)\cdot{}T(p)\cdot{}\Delta{}p} \, .
\end{equation}
In the following two sections, the determination of the detector
acceptance and the calculation of the livetime will be described.

\subsection{Detector acceptance}
\begin{figure}[!h]
\begin{center}
\includegraphics[width=14cm]{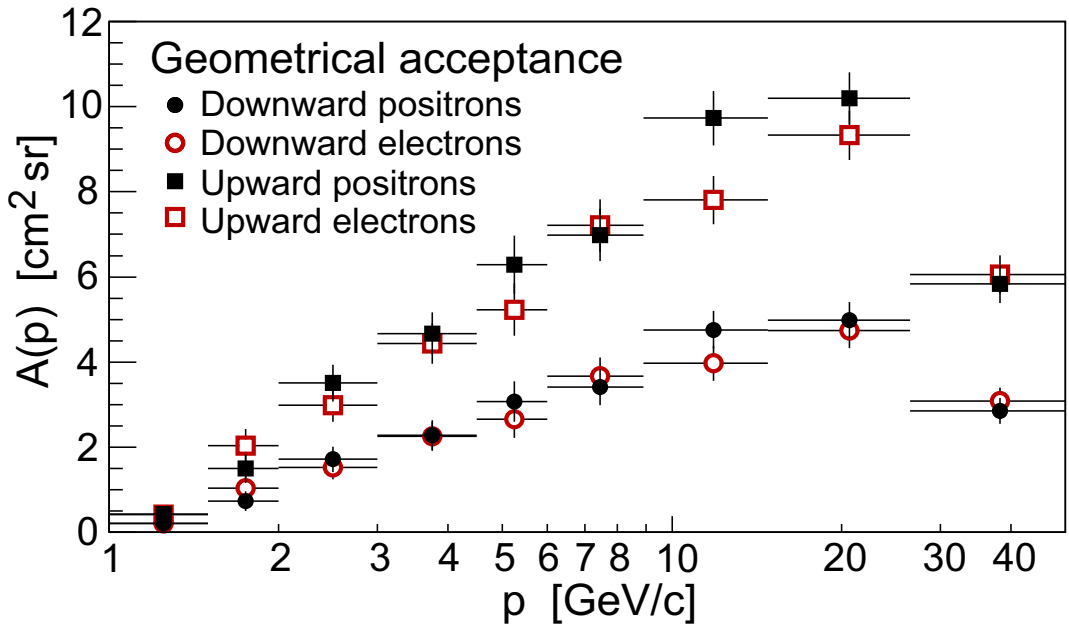}
\caption{\label{fig:acceptance}The geometrical acceptance for
downward (circles) and upward (squares) going positrons (filled) and
electrons (open), when identified through bremsstrahlung conversion.}
\end{center} 
\end{figure}

The detector acceptance for the bremsstrahlung conversion process is
calculated from Monte Carlo, separately for electrons and positrons
and for downward and upward going particles. In the simulation,
particles are emitted from a square surface $S$, with a side length of
3.9\+m, above or below the detector, respectively. With
$n_{\mathrm{t}}$ being the total number of Monte Carlo particles
emitted from $S$ into the hemisphere facing the detector with an
isotropic angular distribution, and $n_{c}$ the number of
reconstructed events remaining after the cuts, the acceptance as a
function of incident momentum is~\cite{sullivan71a}
\begin{equation}
A(p) = S \cdot \pi \cdot \frac{n_{c}(p)}{n_{t}(p)}\, .
\end{equation}
%FIGURE fig:acceptance
As Figure~\ref{fig:acceptance} shows, $A(p)$ is on the order of
several cm$^2\cdot\,$sr and reaches a maximum at approximately
20\+\GeV{}/c. Towards higher momentum the decreasing cluster
separation approaches the resolution limit of the silicon strip
detectors, and the acceptance drops. At low momentum, by contrast,
secondary particles may be deflected such that they generate multiple
separated hits in the TOF scintillators. In this case events are
rejected by the trigger logic of the experiment. Furthermore, the
probability rises that secondary particles have a too low momentum to
be properly reconstructed, hence the acceptance decreases.

Formed by the Space Shuttle's payload bay floor and the support
structure of AMS-01, additional material is traversed by upward going
particles before they enter the detector, thus increasing the
probability of bremsstrahlung emission and
photoconversion. Consequently, the acceptance for upward going
particles is generally higher with respect to downward going ones. The
amount of this additional material is estimated to be 4.5\+\% of a
radiation length. No significant difference in the acceptance for
electrons and positrons is observed.

\subsection{Calculation of livetime}\label{sec:livetime}
Two cardinal effects can prevent cosmic ray particles from reaching
the detector. First, the body of the Earth obstructs particles
arriving from the ``wrong'' side. Second, and more complicated, the
geomagnetic field forces the trajectories of incoming particles on
a helix, effectively capturing particles with under-cutoff
momentum. This effect depends on the position of the Space Shuttle,
the incident direction and time. In addition, the periods of time
during which the trigger system was busy enter as dead time into this
calculation.

The livetime $T(p)$ was derived as follows. The acceptance region of
the AMS-01 detector was divided into nine bins of equal size along
$\cos(\theta)$, in the interval of $[0.7,1]$, and into eight bins
along $\phi$. The momentum range between 1\+\GeV{}/c and 50\+GeV{}/c
was divided into eight bins. Then, for every four seconds during the
flight, using the recorded position and attitude of Discovery and for
each of the 576 $(p,\mathrm{d}\Omega)$ bins, a virtual charged
particle was started with the corresponding values on the aperture of
the detector and propagated backward through the geomagnetic field. If
the virtual particle fulfilled the criteria of a primary cosmic ray
particle as described in
\S~\ref{subsection:cutoff}, the interval during which the trigger was
not busy was added to the total livetime.

\begin{figure}[!h]
\begin{center}
\includegraphics[width=14cm]{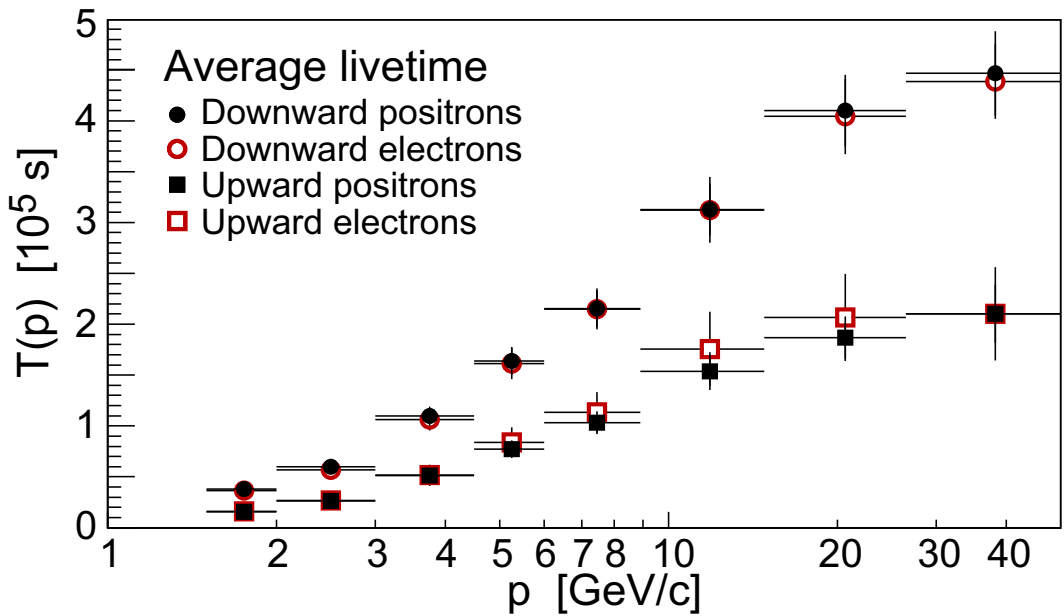}
\caption{\label{fig:livetime}The livetimes, averaged over the detector acceptance, 
as functions of momentum for downward (circles) and upward (squares)
going positrons (filled) and electrons (open).}
\end{center} 
\end{figure}

The livetime, averaged over the detector acceptance, for downward and
upward going positively and negatively charged particles, is displayed
in Figure~\ref{fig:livetime}. Due to obstruction by the Earth, the
livetime for downward going particles is twice that of upward going
ones. Concerning the average livetime, no significant difference
between positively and negatively charged particles is apparent.
%FIGURE fig:livetime

\subsection{Positron and electron fluxes}
\begin{figure}[!h]
\begin{center}
\includegraphics[width=14cm]{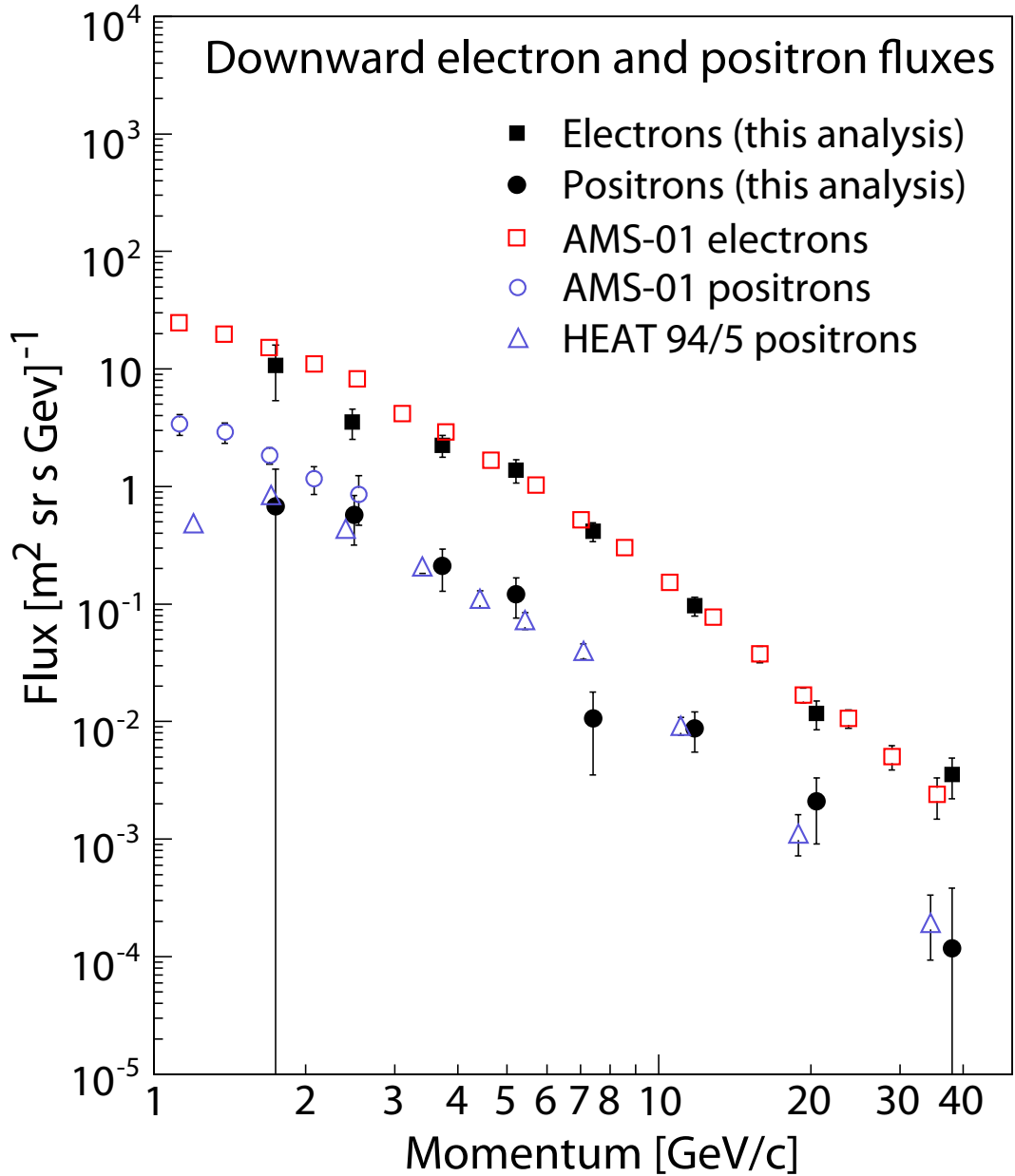}
\caption{\label{fig:fluxes}The fluxes of downward going positrons
(filled circles) and electrons (filled squares) measured in this
analysis, compared with earlier results from AMS-01 (open circles and
squares)~\protect\cite{alcaraz00a} and HEAT-$e^{\pm}$
(triangles)~\protect\cite{duvernois01a}.  Error bars denote
statistical errors only.}
\end{center} 
\end{figure}

Since the amount of material underneath the detector is estimated
only, in this analysis particle fluxes are calculated solely for
particles which traverse the detector
top-down. Figure~\ref{fig:fluxes} displays the fluxes of downward
going positrons and electrons, together with results published earlier
by AMS-01~\cite{alcaraz00a} and HEAT-e$^{\pm}$~\cite{duvernois01a}
with their statistical errors. The fluxes are in very good agreement
with previous measurements over the full momentum range, except for a
slight discrepancy in the electron fluxes between 2 and 3\+\GeV{}/c. Here,
at low momentum in combination with low statistics, we expect the
inaccuracies of the backtracing through the geomagnetic field to
become the dominant source of systematic error to the fluxes. However,
for the positron fraction as a ratio of particle counts, this effect
cancels out.
%FIGURE fig:fluxes

\section{Conclusions}
In this paper, we present a new measurement of the cosmic ray positron
fraction up to energies of 30\+\GeV{} with the AMS-01 detector. Positrons
are identified by conversion of bremsstrahlung photons, which yields
an overall proton rejection on the order of $10^6$. This approach
allows to extend the energy range accessible to the experiment far
beyond its design limits and to fully exhaust the detector's
capabilities. The results, especially on the positron fraction, are
consistent with those obtained in previous experiments at large.

For the reconstruction of converted bremsstrahlung events, customized
algorithms for track finding and event reconstruction have been
developed and implemented. We have shown that the background
is controllable and the overall uncertainty is dominated by the
statistical error due to the low overall cross section of the signal
process. 

Furthermore, the absolute lepton fluxes have been calculated and found
to match the earlier results. This required a new precise and
extensive livetime calculation.

\newpage
\section*{Acknowledgments}
%%%%%%%%%%%%%%%%%%%%%%%%%%%%%%%%%%%%%%%%%%%%%%%%%%%%%%%%%%%%%%%%%%%%%%%%%%%%%%%
%
% copied from 03.
%
The support of INFN, Italy, ETH--Z\"urich, the University of Geneva,
the Chinese Academy of Sciences, Academia Sinica and National Central
University, Taiwan, the RWTH--Aachen, Germany, the University of
Turku, the University of Technology of Helsinki, Finland, the U.S.~DOE
and M.I.T., CIEMAT, Spain, LIP, Portugal and IN2P3, France, is
gratefully acknowledged.

The success of the first AMS mission is due to many individuals and
organizations outside of the collaboration.  The support of NASA was
vital in the inception, development and operation of the experiment.
Support from the Max--Planck Institute for Extraterrestrial Physics,
from the space agencies of Germany (DLR), Italy (ASI), France (CNES)
and China and from CSIST, Taiwan also played important roles in the
success of AMS.
%
%%%%%%%%%%%%%%%%%%%%%%%%%%%%%%%%%%%%%%%%%%%%%%%%%%%%%%%%%%%%%%%%%%%%%%%%%%%%%%%
%\newpage
%
% Bibliography
%%%%%%%%%%%%%%%%%%%%%%%%%%%%%%%%%%%%%%%%%%%%%%%%%%%%%%%%%%%%%%%%%%%%%%%%%%%%%%
%
% Style file to use with mcite.
% 
\bibliographystyle{amsstylem}

%% AUTHOR LIST
%auto-ignore
\typeout{   }     
\typeout{Using author list for AMS paper 06/PhysRep}
\typeout{$Modified: Oct 01 by M. Capell $}
\typeout{!!!!  This should only be used with document option a4p!!!!}
%
%
%
%  L A T E X  version!!
%
%

\newcounter{tutetotcount}
\newcounter{tutetutecount}
\newcounter{tutecount}
\newcounter{namecount}
\newcommand{\tutenum}[1]{%
\stepcounter{tutetotcount}%
\stepcounter{tutecount}%
\ifnum\value{tutecount}=27%
\stepcounter{tutetutecount}%
\addtocounter{tutecount}{-26}%
\fi%
\xdef#1{{\ifnum\value{tutetutecount}>0\alph{tutetutecount}\fi\alph{tutecount}}}}
\def\tute#1{$\,^{#1}$\stepcounter{namecount}}
\newcounter{notecount}
\newcommand{\note}{{\stepcounter{notecount}\thenotecount}}
\tutenum\aachenI           % 1
\tutenum\aachenIII            % 1
\tutenum\lapp              % 4
\tutenum\jhu
\tutenum\lsu
\tutenum\cssa
\tutenum\calt
\tutenum\iee
\tutenum\ihep           % 7
\tutenum\bologna           % 9 
\tutenum\bucharest         % 12
\tutenum\mit               % 14 
\tutenum\ncu               % 48
\tutenum\maryland
\tutenum\korea
\tutenum\florence          % 15
\tutenum\mpi
\tutenum\cern              % 16 
\tutenum\geneva            % 18
\tutenum\grenoble
\tutenum\hefei
\tutenum\hut
\tutenum\ist
\tutenum\lip
\tutenum\csist
\tutenum\madrid            % 24
\tutenum\milan             % 25
\tutenum\kurch
\tutenum\moscow            % 26
\tutenum\perugia           % 31
\tutenum\pisa
\tutenum\ewha           % 31
\tutenum\as
\tutenum\turku
\tutenum\eth               % 46
{
\parskip=0pt
\section*{The AMS--01 Collaboration}
\tolerance=10000
\hbadness=5000
\raggedright
\def\r{\rlap,}
\noindent
M.~Aguilar\r\tute\madrid\
J.~Alcaraz\r\tute\madrid\
J.~Allaby\r\tute\cern\
B.~Alpat\r\tute\perugia\
G.~Ambrosi\r\tute\perugia\
H.~Anderhub\r\tute\eth\
L.~Ao\r\tute\calt\
A.~Arefiev\r\tute\moscow\
P.~Azzarello\r\tute\perugia\
L.~Baldini\r\tute{\bologna,\mit}\
% ...
\typeout{--------------------------------------------------------------}
\typeout{
Imagine that:  <\thenamecount> authors from <\thetutetotcount> institutes.}
\typeout{--------------------------------------------------------------}
\vspace*{-.5\baselineskip}
\rule[.5\baselineskip]{\textwidth}{0.5pt}
\begin{list}{A}{\itemsep=0pt plus 0pt minus 0pt\parsep=0pt plus 0pt minus 0pt
                \topsep=0pt plus 0pt minus 0pt}
\small
\item[$^\aachenI$] I. Physikalisches Institut, RWTH, D-52074 Aachen, Germany$^5$ % NOTE 5, DLR
\item[$^\aachenIII$] III. Physikalisches Institut, RWTH, D-52074 Aachen, Germany$^5$   % NOTE 5, DLR
\item[$^\lapp$] LAPP, Universit\'e de Savoie, CNRS/IN2P3, F-74941 Annecy-le-Vieux Cedex, France
\item[$^\jhu$] Johns Hopkins University, Baltimore, MD 21218, USA
\item[$^\lsu$] Louisiana State University, Baton Rouge, LA 70803, USA
\item[$^\cssa$] Center of Space Science and Application, Chinese Academy of Sciences, 100080 Beijing, China
\item[$^\calt$] Chinese Academy of Launching Vehicle Technology, CALT, 100076 Beijing, China
\item[$^\iee$] Institute of Electrical Engineering, IEE, Chinese Academy of Sciences, 100080 Beijing, China
\item[$^\ihep$] Institute of High Energy Physics, IHEP, Chinese Academy of Sciences, 100039 Beijing, China$^6$ % NOTE 6, NNSF
\item[$^\bologna$] University of Bologna and INFN-Sezione di Bologna, I-40126 Bologna, Italy$^7$  % NOTE 7, ISA
\item[$^\bucharest$] Institute of Microtechnology, Politechnica University of Bucharest and University of Bucharest, R-76900 Bucharest, Romania
\item[$^\mit$] Massachusetts Institute of Technology, Cambridge, MA 02139, USA
\item[$^\ncu$] National Central University, Chung-Li, Taiwan 32054
\item[$^\maryland$] University of Maryland, College Park, MD 20742, USA
\item[$^\korea$] CHEP, Kyungpook National University, 702-701 Daegu, Korea
\item[$^\florence$] CNR--IROE, I-50125 Florence, Italy
\item[$^\mpi$] Max--Planck Institut f\"ur extraterrestrische Physik, D-85740 Garching, Germany
\item[$^\cern$] European Laboratory for Particle Physics, CERN, CH-1211 Geneva 23, Switzerland
\item[$^\geneva$] University of Geneva, CH-1211 Geneva 4, Switzerland
\item[$^\grenoble$] Laboratoire de Physique Subatomique et de Cosmologie, IN2P3/CNRS, F-38026 Grenoble, France
\item[$^\hefei$] Chinese University of Science and Technology, USTC, Hefei, Anhui 230 029, China$^{6}$ % NOTE 6, NNSF
\item[$^\hut$] Helsinki University of Technology, FIN-02540 Kylmala, Finland
\item[$^\ist$] Instituto Superior T\'ecnico, IST, P-1096 Lisboa, Portugal
\item[$^\lip$] Laboratorio de Instrumentacao e Fisica Experimental de Particulas, LIP, P-1000 Lisboa, Portugal
\item[$^\csist$] Chung--Shan Institute of Science and Technology, Lung-Tan, Tao Yuan 325, Taiwan
\item[$^\madrid$] Centro de Investigaciones Energ{\'e}ticas, Medioambientales y Tecnol\'ogicas, CIEMAT, E-28040 Madrid, Spain$^8$ % NOTE 8 CICT
\item[$^\milan$] INFN-Sezione di Milano, I-20133 Milan, Italy$^{7}$ % NOTE 7 ISA
\item[$^\kurch$] Kurchatov Institute, Moscow, 123182 Russia
\item[$^\moscow$] Institute of Theoretical and Experimental Physics, ITEP, Moscow, 117259 Russia
\item[$^\perugia$] INFN-Sezione di Perugia and Universit\`a Degli Studi di Perugia, I-06100 Perugia, Italy$^{7}$ % NOTE 7 ISA
\item[$^\pisa$] INFN-Sezione di Pisa and Universit\`a di Pisa, I-56100 Pisa, Italy$^{7}$ % NOTE 7 ISA
\item[$^\ewha$] Ewha Womens University, 120-750 Seoul, Korea
\item[$^\as$] Institute of Physics, Academia Sinica, Nankang Taipei 11529, Taiwan
\item[$^\turku$] University of Turku, FIN-20014 Turku, Finland
\item[$^\eth$] Eidgen\"ossische Technische Hochschule, ETH Z\"urich, CH-8093 Z\"urich, Switzerland

\setcounter{notecount}{0}
% 2*ionica, dinu
\item[$^1$] Permanent address:  HEPPG, Univ.~of Bucharest, Romania.
% G. Lamanna
\item[$^2$] Present address: LAPP, Universit\'e de Savoie, CNRS/IN2P3, 
	    F-74941 Annecy-le-Vieux Cedex, France

% B. Verlaat
\item[$^3$] Now at National Institute for High Energy Physics, NIKHEF, 
            NL-1009 DB Amsterdam, The Netherlands.
% 
% S.Ye
\item[$^4$] Supported by ETH Z\"urich.
%
% noted in institute list
%
% for Aachen I & III
\item[$^5$]  Supported by the 
Deutsches Zentrum f\"ur Luft-- und Raumfahrt, DLR.
%
% for IHEP & Hefei
\item[$^6$] Supported by the National Natural Science Foundation of China.
%
% for Bologna, Milano & Perugia
\item[$^7$] Also supported by the Italian Space Agency.
%
% for CIEMAT
\item[$^8$] Also supported by the Comisi\'on Interministerial de Ciencia y 
           Tecnolog{\'\i}a.
\end{list}
}

%% THE TABLE
\clearpage
\begin{table}[hb]
\begin{center}
%\begin{tabular}{@{[\,}r@{,\,}rclll}
\begin{tabular}{|r@{\,--}r|r|r@{.}l|c|l|l|l|}
\hline
\multicolumn{2}{|c|}{Momentum} & \LL{$N_{e^{\,-}}$} & \multicolumn{2}{|c|}{\LL{$N_{e^{\,+}}$}} & Positron & \LL{$\sigma_{\mathrm{stat}}$} & \LL{$\sigma_{\mathrm{sys,b}}$} & \LL{$\sigma_{\mathrm{sys,l}}$} \STRUT \\
\multicolumn{2}{|c|}{[\GeV{}/c]} & & \multicolumn{2}{|c|}{} & fraction & & & \\ \hline
~1.0 & ~1.5    & 11   & 3&0   &  0.210   &    $_{-0.1} ^{+0.11}$    &   $\pm 0$               &   $\pm 0$     \STRUT \\
~1.5 & ~2.0    & 31   & 4&8   &  0.133   &    $_{-0.051}^{+0.064}$  &   $_{-0}^{+0.002}$      &   $\pm 0.006$ \STRUT \\
~2.0 & ~3.0    & 85   & 10&7  &  0.112   &    $_{-0.031}^{+0.034}$  &   $_{-0.003}^{+0.001}$  &   $\pm 0.004$ \STRUT \\
~3.0 & ~4.5    & 186  & 15&8  &  0.078   &    $_{-0.018}^{+0.021}$  &   $_{-0.003}^{+0.001}$  &   $\pm 0.004$ \STRUT \\
~4.5 & ~6.0    & 172  & 10&0  &  0.055   &    $_{-0.022}^{+0.025}$  &   $_{-0.007}^{+0.006}$  &   $\pm 0.001$ \STRUT \\
~6.0 & ~8.9    & 198  & 9&0   &  0.043   &    $_{-0.017}^{+0.029}$  &   $_{-0.004}^{+0.01}$   &   $\pm 0.004$ \STRUT \\
~8.9 & 14.8    & 195  & 14&5  &  0.069   &    $_{-0.014}^{+0.03}$   &   $_{-0.002}^{+0.01}$   &   $\pm 0.006$ \STRUT \\
14.8 & 26.5    & 109  & 15&4  &  0.124   &    $_{-0.03}^{+0.038}$   &   $_{-0.003}^{+0.009}$  &   $\pm 0.007$ \STRUT \\
26.5 & 50.0    & 39   & 2&9   &  0.070   &    $_{-0.034}^{+0.075}$  &   $_{-0.01 }^{+0.01}$   &   $\pm 0.007$  \STRUT \\
\hline
\end{tabular}
\caption{\label{table:fraction}The number of electron ($N_{e^{\,-}}$) and corrected positron ($N_{e^{\,+}}$) 
candidates and the positron fraction as a function of momentum.
Systematic errors are given separately for background
($\sigma_{\mathrm{sys,b}}$) and livetime ($\sigma_{\mathrm{sys,l}}$)
correction.}
\end{center}
\end{table}

\end{document}